\documentclass[showpacs,preprintnumbers,superscriptaddressprl, twocolumn,showpacs, floatfix, groupedaddress,prd]{revtex4}
\usepackage{graphicx}
\usepackage{amsfonts}
\usepackage{amsmath, amsthm, amssymb}

\newcommand{\ie}{\textit{i.e. }}
\newcommand{\eg}{\textit{e.g. }}

\begin{document}

\title[Parametric resonance for antineutrino conversions using LSND best-fit results with a 3+1 flavor scheme]{Parametric resonance for antineutrino conversions using LSND best-fit results with a 3+1 flavor scheme}
\author{J. Linder}
\email{jacob.linder@ntnu.no}
\affiliation{Department of Physics, Norwegian University of
Science and Technology, N-7491 Trondheim, Norway}
\date{Received \today}
\begin{abstract}
An analytical solution to a parametric resonance effect for antineutrinos in a 3+1 flavor (active+sterile) scheme using multiple non-adiabatic density shifts is presented. We derive the conditions for a full flavor conversion for antineutrino oscillations $\overline{\nu}_\alpha \to \overline{\nu}_s$ $(\alpha=e,\mu,\tau)$ under the assumption that LSND best-fits for the mixing parameters are valid in a short-baseline accelerator experiment. We show that the parametric resonance effect can be exploited to increase the effective antineutrino oscillation length by a factor of 10-40, thus sustaining a high oscillation probability for a much longer period of time than in the vacuum scenario. We propose a realistic experimental setup that could probe for this effect which leaves a signature in terms of a specific oscillation probability profile. Moreover, since the parametric resonance effect is valid in any 2 or 1+1 flavor approximation, our results could be suggestive for future short-baseline accelerator neutrino detection experiments.
\end{abstract}
\pacs{13.15.+g, 14.60.Lm, 14.60.Pq} 
\maketitle
\section{Introduction}
Ever since Pontecorvo gave birth to the concept of neutrino oscillations in 1958  \cite{ponte}, 
the model has evolved significantly over the years \cite{msw3,msw2}, allowing us to have a 
comprehensive understanding of neutrino behaviour 
today. Oscillations take place in both vacuum and matter. The presence of a medium gives rise to matter potentials felt by the neutrinos propagating through it \cite{wolf,langacker,ms,bethe} which in short alters the 
oscillation probability compared to propagation in vacuum. This effect can have dramatic consequences in resonance situations. For a non-adiabatic density shift, the exact analytical solution for the oscillation probability is available in {\it e.g.} Ref. \cite{not}, while various uses of a single and multiple shifts have been studied in {\it e.g.} Refs. \cite{study1,study2,study3}. The idea of using multiple unit-step density shifts to obtain a full flavor conversion for neutrinos was first studied in Ref. \cite{jacphys}. \\
\indent Although the scenario of neutrinos propagating through several layers of matter with constant densities certainly is not new, this paper contributes to this particular topic through the study of resonance conditions for antineutrino flavor conversions in a 3+1 flavor scenario with LSND best-fit parameters. One could very well ask for a justification concerning the choice of a 3+1 flavor scenario as there certainly has been voices in favor of for instance a 3+2 solution \cite{lsnd2}. Adding one sterile neutrino flavor $\nu_s$ represents the simplest extension of the Standard model, containing an active neutrino flavor for each lepton generation $e,\mu,\tau$. The need for such an extension comes from the discrepancy between solar and atmospheric neutrino detection experimental data \cite{gallex,kamiokande,super-kamiokande,SNO} and the short-baseline accelerator LSND \cite{lsnd} results if one insists on using a 3+0 scheme. Consistency can be regained by adding a sterile neutrino flavor, thus obtaining three independent square mass differences $\Delta m^2$. The opportunity to significantely influence the oscillation probability and oscillation length arises in a 3+1 flavor scenario under valid approximations concerning the square mass differences, as we will show. This the main motivation for selecting the 3+1 flavor scheme. 

\par
This paper is organized as follows. In Sec. \ref{sec:3+1} we establish the 1+1 flavor approximation for LSND best-fit results, while in Sec. \ref{sec:pre} the parametric resonance effect is treated for this scheme. The opportunities that arise from the parametric resonance effect with regard to exercising control over the neutrino oscillation probability and oscillation length are discussed in Sec. \ref{sec:discuss}, where we also show that this effect would be observable in a realistic experimental setup. Concluding remarks are given in Sec. \ref{sec:summary}.

\section{Oscillations in the 3+1 flavor scheme}\label{sec:3+1}
\begin{figure}[h!]
\centering
\resizebox{0.27\textwidth}{!}{
	\includegraphics{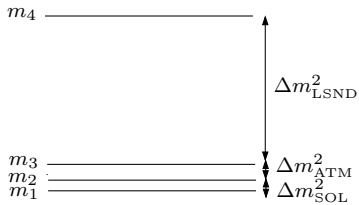}}
\caption{Assumed mass hierachy scheme for neutrino mass eigenstates $|\nu_i(\mathbf{x},t)\rangle$.}
\label{fig:mass1}
\end{figure}

\indent The neutrino flavor states $|\nu_\alpha\rangle$ ($\alpha = e,\mu,\tau,s$) 
are linear superpositions of the mass eigenstates $|\nu_i\rangle$ ($i = 1,2,3,4$) and vice versa. The mixing between these two sets of states explicitely reads
\begin{equation}
|\nu_\alpha(\mathbf{x},t)\rangle = \sum_{i=1}^4 U_{\alpha i}|\nu_i(\mathbf{x},t)\rangle,
\end{equation}
where $U$ is a unitary $4 \times 4$ neutrino mixing matrix which we parametrize in a similar manner as Ref. \cite{gg1}, 
\begin{equation}\label{eq:U}
U = U_{34}U_{24}U_{23}U_{14}U_{13}U_{12}.
\end{equation}
The matrices $U_{ij}$ represent mixing between pairs of mass eigenstates and have the generic form of a rotation with an angle $\theta_{ij}$ in 2-dimensional $i-j$ space. For example, one has
\begin{align}\label{eq:Upara}
U_{23} = \left[ \begin{array}{cccc}
					1 & 0 & 0 & 0 \\
					0 & c_{23} & s_{23} & 0 \\
					0 & -s_{23} & c_{23} & 0 \\
									0 & 0 & 0 & 1 \\
     					\end{array} \right],
\end{align}
where $c_{ij} \equiv \cos \theta_{ij}$, $s_{ij} \equiv \sin \theta_{ij}$. In writing down $U$, we have intentionally left out the CP-violating phase $\delta_{\mbox{\tiny CP}}$, assuming it is sufficiently small to be neglected, and Majorana mass phases, which are irrelevant with respect to oscillations. The time dependence of the mass eigenstates is governed by $|\nu_i({\bf x},t)\rangle = \mbox{e}^{-\mbox{\scriptsize i}E_it} |\nu_i({\bf x},0)\rangle$, where $E_i = \sqrt{\mathbf{p}^2 + m{_i}^2} \simeq E + m{_i}^2/2E$, with the definition $|\mathbf{p}| \equiv E$. Consequently, we operate with neutrino momenta $\mathbf{p}$ much greater than their masses $m_i$, which is an excellent approximation, and assume equal momentum among the mass eigenstates. Furthermore, we adapt the mass hierarchy scheme in Fig. \ref{fig:mass1} with the definitions $\Delta m_{12}^2 = \Delta m_{\mbox{\tiny SOL}}^2$, $\Delta m_{23}^2 = \Delta m_{\mbox{\tiny ATM}}^2$, $\Delta m_{34}^2 = \Delta m_{\mbox{\tiny LSND}}^2$ ($\Delta m_{ij}^2 \equiv |m_i^2 - m_j^2|$).

The neutrino oscillation probability is calculated in standard quantum mechanical fashion $P_{\alpha\to\beta} = |\langle\nu_\beta(\mathbf{x},t) |\nu_\alpha(\mathbf{x},t)\rangle|^2$, producing
\begin{equation}\label{eq:3prob}
P_{\alpha\to\beta} = \sum_{\stackrel{i\in{1,2,3,4}}{_{j\in{1,2,3,4}}}} J_{\alpha\beta ij}\mbox{e}^{-\mbox{\scriptsize i}\Delta m\stackrel{2}{_{ij}}t/2E}
\end{equation}
\noindent with $J_{\alpha\beta ij} = U_{\beta i} U_{i \alpha}^* U_{\beta j}^* U_{j \alpha}$.
The orthonormality condition $\langle \nu_i({\bf x},t)|\nu_j({\bf x},t)\rangle = \delta_{ij}$ is assumed to hold. 

\par
 Considering the mass hierarchy scheme of Fig. \ref{fig:mass1}, it is seen that Eq. (\ref{eq:3prob}) can be significantely simplified. The best-fit results from the LSND experiment states that $\Delta m_{\mbox{\tiny LSND}}^2 \simeq 1.2$ eV$^2$ \cite{lsnd}, such that it is a very good approximation to set $\Delta m_{\mbox{\tiny LSND}}^2 \gg \{\Delta m_{\mbox{\tiny SOL}}^2, \Delta m_{\mbox{\tiny ATM}}^2\}$ since the best-fit equivalents for these are $7.1\times10^{-5}$ eV$^2$ and $2.5\times10^{-3}$ eV$^2$, respectively \cite{gallex,kamiokande,super-kamiokande,SNO}. Upon this approximation, Eq. (\ref{eq:3prob}) turns into (see {\it e.g.} Ref. \cite{lsnd2})
\begin{equation}\label{eq:p1}
P_{\alpha\to\beta} = \delta_{\alpha\beta} - 4U_{\alpha4}U_{\beta4}[\delta_{\alpha\beta} - U_{\alpha4}U_{\beta4}]\sin^2\frac{\Delta m_{a4}^2 t}{4E}.
\end{equation}
Eq. (\ref{eq:p1}) is nothing but the ordinary oscillation formula in the 2-flavor scenario in disguise. One can understand this effective reduction from 3+1 to 1+1 flavors qualitatively by what is referred to as the "quasi two neutrino approximation" or "one mass scale dominance" \cite{pdd1}. For our analysis, this translates into $\Delta m_{\mbox{\tiny LSND}}^2$ dominating the mass scale, thus also dominating the oscillation arena. This justifies an approximative model of two flavors for the LSND short-baseline accelerator experiment. We are immediately able to re-write Eq. (\ref{eq:p1}) as the familiar formula
\begin{equation}\label{eq:p2}
P_{\alpha\to s}= \sin^2(2\theta_{a4})\times\sin^2\Big(\frac{\Delta m_{\mbox{\tiny LSND}}^2 t}{4E}\Big).
\end{equation}
\indent Having gone from 3+1 to 1+1 flavors, an interesting opportunity arises. The parametric resonance effect \cite{jacphys} is now directly applicable in the approximation that $\Delta m_{\mbox{\tiny SOL}}^2$ and $\Delta m_{\mbox{\tiny ATM}}^2$ are negligible compared to $\Delta m_{\mbox{\tiny LSND}}^2$. Due to the large value of $\Delta m_{\mbox{\tiny LSND}}^2$, the corresponding oscillation length is relatively short; of order 10-100 m for neutrinos with $E \sim$ 10-100 MeV. It is important to realize that since $\Delta m_{\mbox{\tiny LSND}}^2 \equiv \Delta m_{34}^2 \simeq$ $\Delta m_{24}^2 \simeq$ $\Delta m_{14}^2$, the following analysis is applicable to the $\overline{\nu}_\alpha \to \overline{\nu}_s$ $(\alpha=e,\mu,\tau)$ channel. The consequences of an enhanced rate of flavor conversion from $\overline{\nu}_\alpha \to \overline{\nu}_s$ could be tested in a disappearance experiment, where a lack of $\overline{\nu}_\alpha$ signatures would indicate the occurence of such oscillations.

\section{Parametric resonance condition}\label{sec:pre}
\begin{figure}[h!]
\centering
\resizebox{0.44\textwidth}{!}{
	\includegraphics{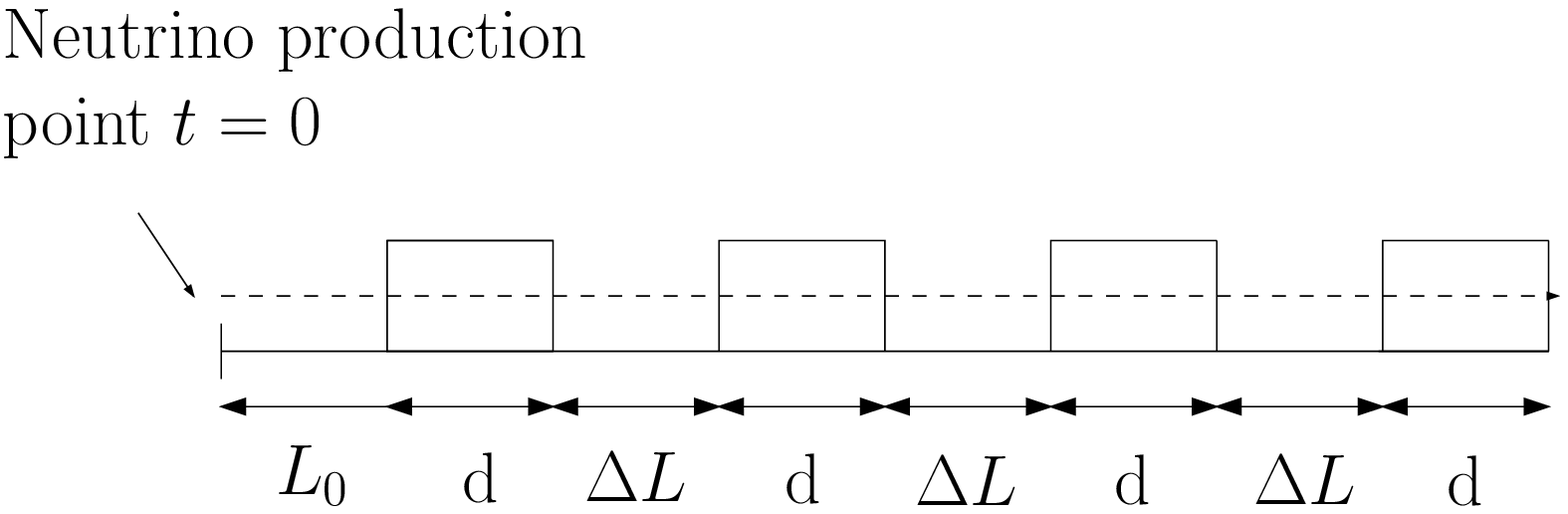}}
\caption{Using multiple non-adiabatic density shifts to obtain a flavor conversion $\overline{\nu}_\alpha \to \overline{\nu}_s$. Material slabs of density $\rho$ and length $d$ are placed at regular intervals in the neutrino trajectory.}
\label{fig:multiunit}
\end{figure}

We now explicitely derive the parametric resonance conditions for a full antineutrino flavor conversion in the $\overline{\nu}_\alpha \to \overline{\nu}_s$ channel. Consider an experimental setup as in Fig. \ref{fig:multiunit}.
Since oscillations are altered in the presence of matter, we must take neutrino forward scattering reactions with 
particles into account. A list of such matter potentials can be found in Ref. \cite{pevsner}. For oscillations between active flavors, it is a known result that only the charged-current (CC) contribution from scattering on electrons is of significance with respect to oscillations since the neutral-current (NC) contribution just gives an overall phase which is irrelevant. In the present case, we consider oscillations between one active and one sterile flavor. The equation of motion for the antineutrinos thus reads
\begin{equation}\label{eq:mot}
\text{i}\frac{\text{d}}{\text{d}t} 
\begin{bmatrix}
\overline{\nu}_\alpha\\
\overline{\nu}_s
\end{bmatrix}
= (H_0 + H_I)
\begin{bmatrix}
\overline{\nu}_\alpha\\
\overline{\nu}_s
\end{bmatrix}
\end{equation}
where $H_I$ contains the forward scattering potentials and is given by
\begin{equation}\label{eq:Hi}
H_I = \frac{1}{2E}\begin{bmatrix}
							  -(V_W^\alpha + V_Z^\alpha) & 0\\
							  0	& 0
							  \end{bmatrix}.
\end{equation}
The CC and NC potentials read $V_W^\alpha = \sqrt{2}G_FN_\alpha$ and $V_Z^\alpha = -G_FN_n/\sqrt{2}$ \cite{wolf}, where $G_F$ is the Fermi constant, while $N_\alpha$ and $N_n$ are the number densities of leptons with flavor $\alpha$ and neutrons in the medium, respectively. The minus sign in Eq. (\ref{eq:Hi}) comes from the fact that the potential sign for antineutrinos is reversed compared to neutrinos \cite{jacmat}. Accordingly, antineutrinos under the influence of a potential must be described by a different set of mixing parameters in the medium than in vacuum. In short, the transformation from the vacuum scenario to the presence of matter is obtained by substituting the vacuum parameters with their matter equivalents, i.e. $\theta \to \theta_m$ and $\Delta m^2 \to \Delta M^2$, for the case of a constant, isotropic electron density. In the present case, these read
\begin{align}\label{eq:mediumparameters}
\Delta M^2 &= \Delta m^2 \sqrt{s_{2\theta}^2 + ({\cal{N}} - c_{2\theta})^2} \notag \\
s_{2\theta_m}^2 &= \frac{s_{2\theta}^2}  {s_{2\theta}^2 + ({\cal{N}} - c_{2\theta})^2},
\end{align}
where the notation 
\begin{equation}\label{eq:CrazyN}
{\cal{N}} = -\frac{2\sqrt{2}G_F(N_\alpha-N_n/2)E}{\Delta m^2}
\end{equation}
 has been introduced. For short-baseline accelerator experiments, one can effectively set $N_\alpha=0$ for $\alpha\in\{\mu,\tau\}$ such that only the NC potential contributes. For $\alpha=e$, however, one must include $N_e$ in Eq. (\ref{eq:CrazyN}).
 
\par
 Consider now the flavor state vector $|\Psi({\bf x},t)\rangle$ describing the antineutrino state. Our initial condition is that $|\Psi({\bf x},0)
\rangle$ = $[1 \mbox{ } 0]^{\mbox{\tiny T}}$, using a basis where 
\begin{align}
|\overline{\nu}_\alpha(\mathbf{x},0)\rangle = [1 \mbox{ } 0]^{\mbox{\tiny T}},\;\;
|\overline{\nu}_s(\mathbf{x},0)\rangle = [0 \mbox{ } 1]^{\mbox{\tiny T}}.
\end{align}
If a non-adiabatic density shift occurs at $t=t_0$, the neutrino propagator that takes the neutrino state vector from $t=0$ to $t=t_0$ reads
\begin{equation}
{\cal{K}}(t) = U\mbox{diag}(\mbox{e}^{-\mbox{\scriptsize i}E_a t}, \mbox{e}^{-\mbox{\scriptsize i}E_4 t})U^{-1}.
\end{equation}
Here, $U$ is a $2\times2$ mixing matrix corresponding to a rotation with an angle $\theta_{as}$ in the mass eigenstate-plane for $\overline{\nu}_\alpha$ and $\overline{\nu}_s$, such that
\begin{equation}\label{eqflavtomass}
			\left[ \begin{array}{c}
			\overline{\nu}_\alpha \\
			\overline{\nu}_s  \\
			\end{array} \right]_{\mbox{\scriptsize flav}} = \left[ \begin{array}{cc}
			c_{a 4} & s_{a 4} \\
			-s_{a4} & c_{a4} \\
			\end{array} \right]
			\left[ \begin{array}{c}
			\nu_a\\
			\nu_4 \\
			\end{array} \right]_{\mbox{\scriptsize mass}}.
\end{equation}
As pointed out in Ref. \cite{msw2}, the flavor eigenstates must be continuous in the transition between vacuum and a 
massive medium even if the density change is extremely non-adiabatic. Demanding that the flavor eigenstates are continuous across the density shift, we write 
$|\Psi({\bf x},t_0)\rangle^{\mbox{\scriptsize mass}}$ = $U_m^{-1}|\Psi({\bf x},t_0)\rangle$, where the subscript $m$ denotes our transition to a massive medium. The propagation through the medium is described by replacing the vacuum quantities with their medium equivalents, such that the flavor state vector in the medium is
\begin{equation}\label{twoprop}
|\Psi({\bf x},t,t_0)\rangle = {\cal{K}}_m(t-t_0){\cal{K}}(t_0)|\Psi({\bf x},0)\rangle, \hspace{0.1in} t\geq t_0.
\end{equation}
\noindent Here, ${\cal{K}}_m(t-t_0)$ is defined in a similar fashion as the vacuum propagator, namely 

\begin{equation}
{\cal{K}}_m(t) = U_m\mbox{diag}(\mbox{e}^{-\mbox{\scriptsize i}E_{am}t}, \mbox{e}^{-\mbox{\scriptsize i}E_{4m}t})U_m^{-1}.
\end{equation}

\begin{figure}
\resizebox{0.5\textwidth}{!}{
	\includegraphics{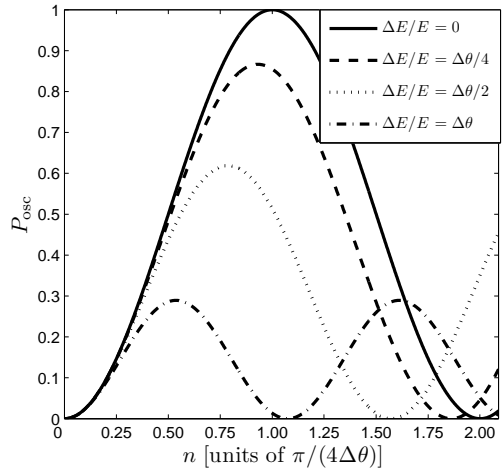}}
\caption{Oscillation probability $P_\text{osc}$ for the setup in Fig. \ref{fig:multiunit} as a function of the number of iterations $n$. The parametrically enhanced oscillation probability reaches unity at $n=\pi/4\Delta\theta$ for $\Delta E$ = 0. }
\label{fig:multipleprob}
\end{figure}

Eq. (\ref{twoprop}) thus gives us the opportunity of using the ${\cal{K}}(t)$ and 
${\cal{K}}_m(t)$ propagators to "rotate" the original flavor eigenstate $\overline{\nu}_\alpha$ into a $\overline{\nu}_s$ by an appropriate choice of parameters. For a setup such as in Fig. \ref{fig:multiunit}, we obtain
\begin{equation}
|\Psi({\bf x},L_0,\Delta L,n,d)\rangle =  {\cal{K}}_{\mbox{\tiny TOT}}(L_0,\Delta L,n,d)|\Psi({\bf x},0)\rangle,
\end{equation}
\begin{widetext}
\begin{align}\label{eq:multiunit2}
{\cal{K}}_{\mbox{\tiny TOT}}(L_0,\Delta L,n,d) &= {\cal{K}}_m(d)\times\Big[{\cal{K}}(\Delta L){\cal{K}}_m(d)\Big]^{(n-1)}{\cal{K}}(L_0), \;\; n=1,2,3,..
\end{align}
Now,  Eq. (\ref{eq:multiunit2}) can be written
\begin{align}\label{eq:psi2}
|\Psi({\bf x},L_0,\Delta L,n,d)\rangle &= [U_mD_m(d)U_m^{-1}UD(\Delta L)U^{-1}]^{(n-1)} U_mD_m(d)U_m^{-1}UD(L_0)U^{-1}|\Psi({\bf x},0)\rangle,
\end{align}
\noindent where $D(x)$ = diag$(1,\mbox{e}^{-\mbox{\scriptsize i}\Delta m_{\mbox{\tiny LSND}}^2x/2E})$
and $D_m(x)$ = diag$(1,\mbox{e}^{-\mbox{\scriptsize i}\Delta M^2x/2E})$. To obtain the diagonal matrices $D$ and $D_m$, we have extracted common phase factors of the type e$^{\mbox{\scriptsize i}\varphi}$ which are irrelevant when pursuing the oscillation probability. The resonance $|\Psi({\bf x},L_0,\Delta L,n,d)\rangle$ = $[0 \mbox{ } 1]^{\mbox{\tiny T}}$ is obtained when we set $L_0 = \Delta L$ and choose the phases such that 
\begin{align}\label{eq:phascond}
\Delta m_{\mbox{\tiny LSND}}^2\Delta L/2E = \pi,\; \Delta M^2d/2E = \pi. 
\end{align}
\noindent As a consequence, $D(\Delta L)$ = $D_m(d)$ = diag$(1,-1)$, and one obtains
\begin{align}
[U_mD_m(D)U_m^{-1}]&[UD(\Delta L)U^{-1}]  = \left[ \begin{array}{cc} \cos 2(\theta_m-\theta_{a4}) & \sin 2(\theta_m-\theta_{a4}) \\ -\sin 2(\theta_m-\theta_{a4})&\cos 2(\theta_m-\theta_{a4})\\ \end{array} \right]  = \mbox{e}^{2\mbox{\scriptsize i}(\theta_m - \theta_{a4})\sigma^y},
\end{align}
\end{widetext}
\noindent where $\sigma^y$ is the Pauli matrix
\begin{equation}
\sigma^y = \left[ \begin{array}{cc} 0 & -\mbox{i} \\ \mbox{i} & 0\\ \end{array} \right]. 
\end{equation}
\noindent Eq. (\ref{eq:psi2}) then becomes $|\Psi({\bf x},L_0,\Delta L,n,d)\rangle$ = $\mbox{e}^{2\mbox{\scriptsize i}n(\theta_m - \theta_{a4})\sigma^y}$$|\Psi({\bf x},0)\rangle$, written out as 
\begin{equation}\label{eq:resprob}
|\Psi({\bf x},L_0,\Delta L,n,d)\rangle = \left[ \begin{array}{c} \cos (2n\Delta\theta) \\ -\sin (2n\Delta\theta) \\ \end{array}\right],
\end{equation}
\noindent with the definition $\Delta\theta \equiv (\theta_m - \theta_{a4})$. The resonance condition for the number of 
iterations $n$ is then 
\begin{equation}\label{eq:resonanceeq}
2n\Delta\theta = \pm\pi/2.
\end{equation}
If this condition is met, the probability for a flavor conversion 
reads $P_{\alpha s} = \sin^2(2n\Delta\theta) = 1$. It is important to realize that this resonance crucially 
depends on choosing the phases properly. To see this, recall that $d$ and $\Delta L$ must be chosen to satisfy Eq. (\ref{eq:phascond}). If $d$ and $L$ instead are slightly perturbed to satisfy 
\begin{align}\label{eq:perturb}
\Delta m_{\mbox{\tiny LSND}}^2\Delta L/2(E+\Delta E) = \pi \notag \\
\Delta M^2d/2(E+\Delta E) = \pi,
\end{align}
the conversion probability never reaches unity and is displaced. This is shown in Fig. \ref{fig:multipleprob}.
\noindent In general, the width of the resonance depends on $\Delta\theta$. For $\Delta E/E \sim \Delta\theta$, it is seen from the graph that one still gets reasonably close to the resonance. Larger values of $\Delta E$ destroys the resonance, and gives a flat curve for the oscillation probability. \\

\section{Discussion of results}\label{sec:discuss}
We showed above that it is, in theory, possible to obtain a full conversion of  $\overline{\nu}_\alpha \to \overline{\nu}_s$ $(\alpha=e,\mu,\tau)$ in a short-baseline accelerator experiment with LSND best-fits for the neutrino mixing variables {\it given} a resonance set of experimental parameters. While the required material slab lengths are manageable, there is also the issue of material density. When restricted to realistic densities in the range $\rho = [1,11]$ g/cm$^3$, 
one obtains $\Delta\theta \sim 10^{-7}$ for the LSND best-fit $\theta_{\mbox{\tiny LSND}} \simeq 0.019$, resulting in an extremely narrow resonance. The number of iterations required 
to reach the resonance of $P_{\alpha s}$ will thus be very high unless there is a way of obtaining a large $\Delta\theta$ in the medium. 

\par 
Although the idea of obtaining the full resonance for antineutrino conversions is unrealistic for material densities available in an experimental setup, an interesting opportunity arises from our analysis. For vacuum and matter oscillations in a single material, the oscillation length for an $E=50$ MeV neutrino using LSND parameters is approximately 50 m. We now show that one is able to prolong the \textit{effective} oscillation length to several hundred meters by using the parametric resonance effect in an experimental setup with material slabs of realistic density and length. In this context, the effective oscillation length refers to the distance upon which the resulting oscillation probability pattern completes a full period. Note that this is neither the vacuum oscillation length $l_\text{vac}$ nor its matter equivalent $l_m$, which of course remain unaltered. Accordingly, the parametric resonance effect would constitute a way of manipulating the neutrinos to maintain a large oscillation probability for a longer period of time. Such an effect could be probed for with aim to investigate the validity of the 1+1 flavor approximation for LSND parameters, and could also be suggestive in future neutrino detection experiments using short-baseline accelerators.

\par
\begin{figure}
\resizebox{0.48\textwidth}{!}{
\includegraphics{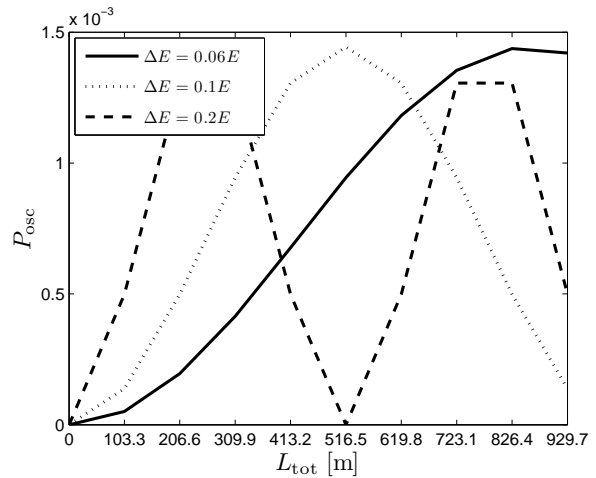}}
\caption{Parametrically enhanced probability for $\overline{\nu}_e\leftrightarrow\overline{\nu}_s$ oscillations using $E=50$ MeV LSND antineutrinos propagating through material slabs with density $\rho = 3$ g/cm$^3$. The oscillation length is significantely increased compared to vacuum oscillations and matter oscillations in a single material slab. For $\overline{\nu}_\alpha\leftrightarrow\overline{\nu}_s$, $\alpha\in\{\mu,\tau\}$ oscillations, these numerical results are also valid since $\Delta\theta$ remains practically unchanged. With a total neutrino trajectory of less than 1 km, a realistic experimental setup would be able to probe for this effect.}
\label{fig:plot}
\end{figure}

Consider Fig. \ref{fig:plot} for numerical simulations of the parametrically enhanced oscillation probability for $\overline{\nu}_e\leftrightarrow\overline{\nu}_s$ oscillations. Several plots are included for different choices of $d$ and $L$, \textit{i.e.} deviations $\Delta E$ in Eq. (\ref{eq:perturb}).  Note that this effect does not occur only within a narrow window of deviations $\Delta E/E\sim \Delta\theta$ which was the condition for the full resonance. Here, deviations as large as 20\% would still cause the oscillation length to be several times larger than in the vacuum scenario. For densities $\rho\in[1,11]$ g/cm$^3$, the numerical results in Fig. \ref{fig:plot} are essentially unchanged. As a consequence, the results are also equally valid for $\overline{\nu}_\mu\leftrightarrow\overline{\nu}_s$ and $\overline{\nu}_\tau\leftrightarrow\overline{\nu}_s$ oscillations since the effect of including $N_e$ in Eq. (\ref{eq:CrazyN}) is negligible, \ie $\Delta\theta$ remains practically unaltered.

\par
Although the maximum oscillation probability in Fig. \ref{fig:plot} is the same as for vacuum oscillations, \ie $P_\text{max} = \sin^2 2\theta_\text{LSND} \simeq 0.0015$, it is clear that the effective oscillation length increases by a factor of 10-40 when using the setup of Fig. \ref{fig:multiunit} with realistic material slab densities. In comparison, neutrino propagation through a single material slab of density $\rho = 3$ g/cm$^3$ has virtually no effect on the oscillation length due to the smallness of $\Delta\theta$. In order to probe for this parametric resonance effect, one would consequently need an experimental setup with consecutive material slabs of density $\rho\sim3$ g/cm$^3$ and an effective neutrino trajectory of less than 1 km. We stress the fact that the idea of substantially increasing the oscillation length by means of the parametric resonance effect is experimentally realistic and constitutes an interesting meta-effect that could yield information concerning the validity of a 2 or 1+1 flavor approximation in a neutrino factory. Thus, the results presented in this paper could prove useful in light of future short-baseline accelerator neutrino detection experiments. 
\par
Finally, we comment upon potential obstacles and competing effects that are vital to take into account when considering this type of parametric resonance effect. For instance, in a disappearance experiment where a deficit of $\bar{\nu}_\alpha$ signatures is measured, it is important to estimate the magnitude of $\bar{\nu}_\alpha\to\bar{\nu}_\beta$ conversions where $\beta\neq s$ in order to be certain of that one is truly observing the present parametric resonance effect. Consider for instance production of $\bar{\nu}_\mu$ at a neutrino source in a disappearance experiment where $\bar{\nu}_\mu\to\bar{\nu}_e$ oscillations would be a competing effect with the parametric resonance presently discussed. Making use of Eq. (\ref{eq:3prob}) to estimate the conversion of active-active type as opposed to
active-sterile, it is found that for the established large mixing angle (LMA) parameters aforementioned for distances $L \simeq 1$ km, $P_{\bar{\nu}_\mu\to\bar{\nu}_e}$ = $5.5\times 10^{-4}$ when the neutrino energy is $E=50$ MeV. This amounts to roughly one third of the maximun oscillation probability in Fig. \ref{fig:plot}, which obviously is a non-negligible contribution to the total conversion rate of $\bar{\nu}_\mu$. Discrimination of active-active conversions from active-sterile conversions could be obtained from \eg the process
$\bar{\nu}_e + p\to e^{+} + n$ in a scintillator detector. We stress that the active-active flavor oscillations are not subject to the parametric resonance conditions given our present values of $L_0, d,$ and $\Delta L$ (see previous section) and simply mimic the ordinary vacuum oscillation probability in spite of the presence of slabs. The active-sterile flavor oscillations, however, are subject to the interesting manipulation illustrated in Fig. \ref{fig:plot}.
Moreover, it is clear that the observation of this effect demands that the specific oscillation probability profile of Fig. \ref{fig:plot} is identified with enough accuracy to unquestionably distinguish it from pure vacuum oscillations. This would preferably require several detectors placed along the trajectory of the antineutrinos instead of one; \eg an intermediate detector at for instance $L\simeq0.5$ km. Clearly, the solid line shown in Fig. \ref{fig:plot} offers the most promising scenario for the study of the oscillation probability profile since fewer oscillations will have taken place than in the case of the dashed and dotted lines at any distance. Another issue that also deserves comment concerns the  collimation of the neutrino beam. A large angular spreading of the neutrino trajectories could endanger the validity of the parametric resonance conditions for a setup as shown in Fig. \ref{fig:multiunit}. Short-baseline experiments of type LSND with a distance source-detector of less than, say, 100 m, will benefit from a better collimation than distances approaching 1 km. The collimation in the latter case would also be worse than neutrinos with energies of order ${\cal{O}}$(GeV) in long-baseline experiments. As an estimate for the loss of neutrino flux poor collimation can give rise to, thus lowering the number of neutrino trajectories originating from the source that are subject to the parametric resonance conditions, consider the following scenario. Assume that the slabs in Fig. \ref{fig:multiunit} are of height 2 m and that the entire neutrino flux is incident on the first slab located at $L_0$ but spreads out in a cone from the source. At a distance of $L=1$ km, the neutrino flux through the last slab (\eg a detector) would then have been reduced to a fraction $f = 1/R_0^2$, where $R_0$ is the radius of the cross-section of the cone at that distance. With the given conditions, one easily finds that $R_0 =L\tan\phi$ where $\phi = \text{atan}(1 \text{ m}/L_0)$ is the angle at which the neutrino beam is spread compared to horisontal ground. Thus, $f\simeq1/400$ and it is seen that a considerable fraction of the neutrino flux is removed from the trajectories passing through the slabs.
Consequently, this would constitute a possible limiting factor of the applicability of the parametric resonance effect depending on the size of the neutrino flux of the source, or at least place high demands upon the collimation of the neutrino beam.

\section{Summary}\label{sec:summary}
\par In summary, we have derived an exact analytical expression for parametric resonance for antineutrino flavor conversions in a 3+1 flavor scenario for LSND best-fit results. In order to exploit the idea of multiple non-adiabatic density shifts to obtain an oscillation probability that \textit{exceeds} the maximum oscillation probability in vacuum (matter), it follows from Eq. (\ref{eq:resprob}) that 
\begin{equation}
n > \theta_{(m)}/\Delta\theta
\end{equation}
must be satisfied. Consequently, the observability and application of a parametrically enhanced oscillation probability depends on obtaining a large $\Delta\theta$, which is impossible to achieve for realistic densities. However, the parametric resonance effect offers an interesting possibility to increase the \textit{effective} neutrino oscillation length, thus leading to the maximum vacuum oscillation probability being sustained over a much larger interval than in the scenario of pure vacuum oscillations or even oscillations in a single medium. We showed that this effect would be observable in a realistic experimental setup using material slabs of density $\rho \sim 3$ g/cm$^3$ and a total neutrino trajectory of less than 1 km, giving rise to a specific oscillation probability profile where the effective oscillation length is manifested as shown in Fig. \ref{fig:plot}. We stress that the maximum oscillation probability is \textit{not} enhanced over this distance, but rather the effective oscillation length. With regard to detection, a disappearance
experiment could search for $\overline{\nu}_e\to\overline{\nu}_s$ oscillations in both the CC and NC channel,
while $\overline{\nu}_\mu\to\overline{\nu}_s$ and $\overline{\nu}_\tau\to\overline{\nu}_s$ oscillations would be harder to detect since interactions with $\overline{\nu}_i$, $i\in\mu,\tau$ only occur in the NC channel. Furthermore, there are experimental complications associated with probing of the parametric resonance effect in terms of competing active-active flavor oscillations and the collimation of the neutrino beam over a longer distance source-detector (1 km) than most short-baseline accelerator experiments. These arguments must be taken into account upon considering realistic approaches in terms of exploiting the parametric resonance effect for active-sterile antineutrino conversions. It is clear that this parametric resonance of the oscillation probability would be applicable to any 2 or 1+1 flavor scenario. In light of the expected MiniBooNE results, the validity of such an approximation could yield an interesting opportunity to take advantage of this effect. 

\section*{Acknowledgments}
J. L. acknowledges K. Olaussen, E. Akhmedov and M. Kachelrie\ss\mbox{ }for stimulating discussions. K. B\o rkje is thanked for useful comments.


\begin{thebibliography}{99}
\bibitem{ponte} B. Pontecorvo, Sov. Phys. JETP 7, 172 (1958). 
\bibitem{msw3} M. C. Gonzalez-Garcia, Yosef Nir, Rev. Mod. Phys. {\bf 75}, p. 345-402 (2003).
\bibitem{msw2} T. K. Kuo, J. Pantaleone, Rev. Mod. Phys {\bf 61}, p.941-958 (1989).
\bibitem{wolf} L. Wolfenstein, Phys. Rev. D {\bf 17}, p. 2369-2374 (1978).
\bibitem{langacker} P. Langacker, J. P. Leveille, J. Sheiman, Phys. Rev. D {\bf 27}, p. 1228-1242 (1983).
\bibitem{ms} S. P. Mikheyev, A. Y. Smirnov, in Proceedings of the Tenth International Workshop on Weak Interactions, Savonlinna, Finland, 16-25 June 1985 (unpublished)
\bibitem{bethe} H. A. Bethe, Phys. Rev. Lett. {\bf 56}, p. 1305-1308 (1986).
\bibitem{not} D. N\"otzold, Phys. Rev. D {\bf 36}, p. 1625-1633 (1987)
\bibitem{study1} E. K. Akhmedov, Nucl. Phys. B {\bf 538} p. 25-51 (1999).
\bibitem{study2} E. K. Akhmedov, A. Y. Smirnov, Phys. Rev. Lett. {\bf 85} p. 3978 (2000).
\bibitem{study3} M. V. Chizhov, S. T. Petcov, Phys. Rev. Lett. {\bf 83}, p. 1096-1099 (1999).
\bibitem{jacphys} E. K. Akhmedov, Pramana 54, p. 47-63 (2000).
\bibitem{lsnd2} M. Sorel, J. M. Conrad,  M. H. Shaevitz, Phys. Rev. D {\bf 70}, 073004 (2004).
\bibitem{gg1} C. Giunti, M. C. Gonzalez-Garcia, C. Pe\~na-Garay, Phys. Rev. D {\bf 62} 013005 (2000).
\bibitem{gallex} J. N. Bahcall, M. H. Pinsonneault, Rev. Mod. Phys {\bf 64}, p.885-926 (1992); P. Anselmann {\it et al.} (GALLEX collaboration), Phys. Lett. B {\bf 342} 440-450 (1995).
\bibitem{kamiokande} Y. Fukuda {\it et al.}, (The Kamiokande collaboration), Phys. Rev. Lett. {\bf 77}, 1683 (1996). 
\bibitem{super-kamiokande} M. B. Smy {\it et al.},(Super-Kamiokande Collaboration), Phys. Rev. D {\bf 69}, 011104 (2004); S. Fukuda {\it et al.}, (Super-Kamiokande Collaboration), Phys. Rev. Lett. {\bf 86}, 5651 (2001).
\bibitem{SNO} Q. R. Ahmad {\it et al.} (SNO Collaboration), Phys. Rev. Lett. {\bf 87}, 071301 (2001); Q. R. Ahmad {\it et. al} (SNO collaboration), Rev. Mod. Lett {\bf 87}, 071301 (2001). 
\bibitem{lsnd} C. Athanassopoulos {\it et. al}, Phys. Rev. Lett. 81, p. 1774-1777 (1998).
\bibitem{pevsner} W. C. Kim, A. Pevsner, 1993, {\it Neutrinos in Physics and
Astrophysics}, Contemporary Concepts in Physics, No. 8 (Harwood
Academic, Chur, Switzerland) (1993).
\bibitem{jacmat} J. Linder, hep-ph/0504264 (2005).
\bibitem{bugey} B. Achkar {\it et al.}, Nucl. Phys. B {\bf 424}, 503 (1995).
\bibitem{pdd1} K. Hagiwara {\it et al.}, Phys. Rev. D {\bf 66}, 010001 (2002).
\end{thebibliography}
\end{document}